# Controlling magnetic damping with spintronic thermal effects in Py/Fe$_3$O$_4$-PANI bilayers

José Laurentino[1,2], Carlos Eduardo[2,4], Luiza Paffer[2,3], José Araújo[2,3], and José Holanda[1,2,3,4,5*]

[1]Programa de Pós-Graduação em Engenharia Física, Universidade Federal Rural de Pernambuco, 54518-430, Cabo de Santo Agostinho, Pernambuco, Brazil
[2]Group of Optoelectronics and Spintronics, Universidade Federal Rural de Pernambuco, 54518-430, Cabo de Santo Agostinho, Pernambuco, Brazil
[3]Unidade Acadêmica do Cabo de Santo Agostinho, Universidade Federal Rural de Pernambuco, 54518-430, Cabo de Santo Agostinho, Pernambuco, Brazil
[4]Programa de Pós-Graduação em Física Aplicada, Universidade Federal Rural de Pernambuco, 52171-900, Recife, Pernambuco, Brazil
[5]Programa de Pós Graduação em Tecnologias Energéticas e Nucleares (Proten), Universidade Federal de Pernambuco, Recife, 50740-545, PE, Brazil

## ABSTRACT

We report experiments that control magnetic damping in Py/Fe$_3$O$_4$-PANI via two spintronic thermal effects: spin Seebeck and anomalous Nernst. Magnetic damping is measured using ferromagnetic resonance (FMR) techniques, where the sample is excited by microwave radiation and the resulting DC voltage is detected in the Fe$_3$O$_4$-PANI film. When a temperature gradient is applied in the longitudinal configuration, the DC voltage linewidth increases or decreases depending on the direction of the thermal gradient. This finding demonstrates that magnetic damping in Py/Fe$_3$O$_4$-PANI can be controlled by currents induced by thermal effects. Remarkably, the absolute linewidth changes observed in Py/Fe$_3$O$_4$-PANI exceed those seen in bilayers of ferrimagnetic insulator yttrium iron garnet and heavy metal by more than an order of magnitude. We interpret the significant change in magnetic damping as a result of intrinsic phenomena occurring in bilayers.

*Corresponding author: joseholanda.silvajunior@ufrpe.br

# INTRODUCTION

Recent discoveries of new processes for generating, transporting, and detecting spin currents have propelled the field of spintronics toward previously unimagined possibilities [1-7]. Many spintronic phenomena observed in magnetic bilayers arise from the flow of spin angular momentum, which can occur with or without a concomitant flow of charge carriers between ferromagnetic materials (FM) and adjacent materials. Notably, some phenomena studied over the last decade have been observed in single ferromagnetic layers [8, 9-11]. When dynamic properties in ferromagnetic/heavy metal (HM) bilayer are excited by microwaves, two notable effects emerge: spin rectification (SR) and spin pumping (SP). Spin rectification refers to the generation of a direct current (DC) voltage due to the nonlinear coupling between resistance and current driven by magnetization dynamics [4, 12-17]. In contrast, spin pumping is the direct conversion of a spin current into a charge current via spin-orbit interaction [9, 11]. Both SR and SP are associated with ferromagnetic resonance and can be studied simultaneously. These phenomena are particularly significant because they facilitate the conversion between spin currents and charge currents.

From a fundamental perspective, two key phenomena generate spin-charge currents from a thermal gradient: the spin Seebeck effect (SSE) and the anomalous Nernst effect (ANE). The spin Seebeck effect describes the generation of a spin current in ferromagnetic materials (FMs) in response to a thermal gradient. In SSE, when a temperature gradient ($\nabla T$) is applied perpendicularly to the plane of the FM, a spin current flows in the same direction. This spin current can then enter an adjacent normal metal (NM), where it is transformed into a charge current by the inverse spin Hall effect (ISHE) and can be detected as a relevant DC voltage [18-20]. The electric field generated in the NM as a result of the thermal gradient can be expressed as $\vec{E}_{SSE} = -\alpha_S \hat{\sigma} \times \nabla T$, where $\alpha_S$ is the spin-Seebeck coefficient and $\hat{\sigma}$ is represents the spin polarization aligned with the applied magnetic field. In a bilayer comprising a metallic ferromagnetic film and an NM layer with a thermal gradient applied perpendicularly, one would also expect to observe a DC voltage due to the SSE [1, 2]. Additionally, in a single FM film, an electric field is also generated by the classical anomalous Nernst effect, represented as $\vec{E}_{ANE} = -\alpha_N \hat{\sigma} \times \nabla T$, where $\alpha_N$ is the anomalous Nernst coefficient [1-6, 18-28]. In effect, both currents due to SSE and ANE coexist in the FM/HM bilayers, resulting in a single DC voltage.

Traditionally, SR, SP, ANE, and SSE are experimentally studied using different setups and often observed separately. In this study, we investigated the control of magnetic damping through spintronic thermal effects in Py/Fe$_3$O$_4$-PANI bilayers, varying the thickness of the Py layer while maintaining a constant thickness of 200 nm for the Fe$_3$O$_4$-PANI film. We excited the magnetization dynamics using microwave radiation under ferromagnetic resonance conditions, leading to a DC voltage resulting from SR and SP. The control of magnetic damping was influenced by the currents generated from SSE and ANE. This finding represents a significant advancement in the field of spin caloritronics using Py/Fe$_3$O$_4$-PANI bilayers.

**EXPERIMENTAL SETUP**

The Py layers were grown by DC sputtering on (001) Si substrates, which had a rectangular shape with lateral dimensions of 2 mm by 4.5 mm and thicknesses ($t_{Py}$) varying from 5 nm to 100 nm. In contrast, the Fe$_3$O$_4$-PANI layer of 200 nm was deposited using spin coating at a rotation speed of 1000 rpm for first stage and 5000 rpm for second stage. The time between stages was of 30s. The Fe$_3$O$_4$ was mixed with PANI prior to the spin coating process. For the measurement of DC voltage associated with the SR, SP, SSE, and ANE phenomena, two silver strips, each 300 μm wide, were deposited along the edges of the Fe$_3$O$_4$-PANI layer. Thin copper wires were attached to these strips using silver paint and connected to a nanovoltmeter to measure the DC voltage (V). The microwave signal, utilized for the SR and SP measurements, was provided by a generator with a frequency tunable in the range of 5–8 GHz, maintaining a fixed microwave power of $P_S$ = 20 mW. This microwave signal was fed into a microstrip line with a characteristic impedance of 50Ω, which was constructed on a Duroid plate with a copper line 0.8 mm wide, and a ground plane affixed to a copper support. The Py/Fe$_3$O$_4$-PANI sample was placed atop the microstrip and separated by an 80 μm thick Mylar sheet, excited by an rf magnetic field ($h_{rf}$) that was perpendicular to the magnetic field (H) applied in the bilayer plane. The DC voltage (V) generated under ferromagnetic resonance conditions, at a fixed frequency, was measured by sweeping the magnetic field across the complete series of Py/Fe$_3$O$_4$-PANI samples.

To apply a thermal gradient across the bilayer, a commercial Peltier module, 2 mm wide, was utilized to heat or cool the metallic layer, while the Si substrate remained in thermal contact with the copper microstrip and the Duroid plate. The temperature difference (ΔT) across the Py/Fe$_3$O$_4$-PANI sample was calibrated as a function of the current in the Peltier module, using a differential thermocouple. One junction of the thermocouple was attached to a thin copper strip located between the Peltier module and the sample structure, while the other junction was connected to the copper microstrip. After calibration, the thin copper strip and thermocouple were removed to avoid interfering with the DC voltage (V). **Figure 1(a)** illustrates the experimental setup and sample arrangement used for the simultaneous measurements of the SR, SP, SSE, and ANE phenomena in Py/Fe$_3$O$_4$-PANI. The polarity of the measured voltages is indicated by plus and minus signs. **Figure 1(b)** shows the measured resonance field as a function of frequency represented by symbols, while the solid curve was obtained using the Kittel equation [29], $f = \gamma[(H_R + H_A)(H_R + H_A + 4\pi M_{Eff})]^{1/2}$, where $\gamma = 2.8$ GHz/kOe, $4\pi M_{Eff} = 4\pi M_S - H_S = 10.7$ kG, and plane anisotropy field $H_A = 15$ Oe, are the parameters pertinent for Py/Fe$_3$O$_4$-PANI obtained from the best fit. Here, $M_S$ represents the saturation magnetization, $H_S = 2K_S/(M_S t_{Py})$ is the surface anisotropy field, and $K_S$ is the surface anisotropy constant of Py [30]. The agreement between the Kittel equation and the data confirms that the observed voltage signals correspond to the ferromagnetic resonance (FMR) absorption.

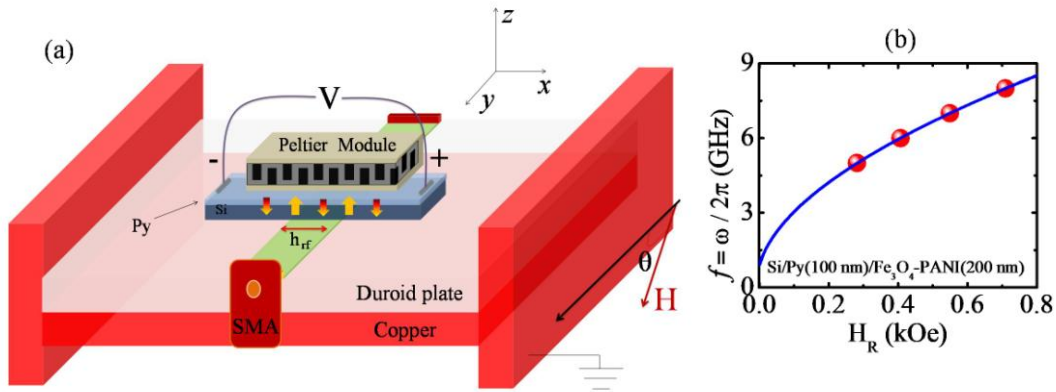

**Figure 1.** (Color online) **(a)** A diagram of the experimental setup and sample arrangement used for simultaneous measurements of the spin rectification (SR), spin pumping (SP), spin Seebeck effect (SSE), and anomalous Nernst effect (ANE) phenomena in Si/Py(100 nm)/Fe$_3$O$_4$-PANI(200 nm). The polarity of the measured voltages is indicated by the plus and minus signs. **(b)** The symbols represent the measured resonance field as a function of frequency, while the solid curve is derived from the Kittel equation [29].

**RESULTS AND DISCUSSIONS**

**Figures 2(a)-(c)** display the measurements of the DC voltage obtained at a microwave frequency of 7 GHz in Py/Fe$_3$O$_4$-PANI bilayers, where the thickness of the Py layer varies between 5 and 30 nm. The magnetic field H is applied along the y-direction ($\theta = 0°$), as illustrated in **Fig. 1(a)**. The observed asymmetric line shape is characteristic of the Spin Relaxation (SR) and Spin Pumping (SP) effects in metallic ferromagnetic (FM)/heavy metal (HM) bilayers. **Figures 2(b)-(d)** present similar measurements taken with a temperature difference of ΔT = -6 K. To comprehend the behavior of the DC voltage generated in Si/Py/$Fe_3O_4$-PANI bilayers due to a thermal gradient, we examined the linewidth derived from the DC voltage peaks. The field-scan voltages exhibit an asymmetric shape that can be fitted using a combination of symmetric Lorentzian and antisymmetric Lorentzian derivative functions, represented by the equation [11]:

$$V(H) = V_{Sym} \frac{(\Delta H)^2}{[(H-H_R)^2+(\Delta H)^2]} + V_{Asym} \frac{\Delta H(H-H_R)}{[(H-H_R)^2+(\Delta H)^2]} . \quad (1)$$

Here, $V_{Sym}$ and $V_{Asym}$ denote the amplitudes of the symmetric and antisymmetric components, while ΔH and $H_R$ represent the ferromagnetic resonance (FMR) linewidth and resonance field, respectively. From the data fits of **Figures 2(a)**, **(b)**, **(c)**, and **(d)** using equation (1), the following linewidths were obtained: 33.65 Oe, 40.82 Oe, 16.22 Oe, and 16.71 Oe for each figure respectively, as indicated by the symmetric components in **Figures 2(e)**, **(f)**, **(g)**, and **(h)**. These results suggest that the linewidth can be modulated by the thermal gradient and serve as evidence of Onsager reciprocity relations between different phenomena in metallic bilayers.

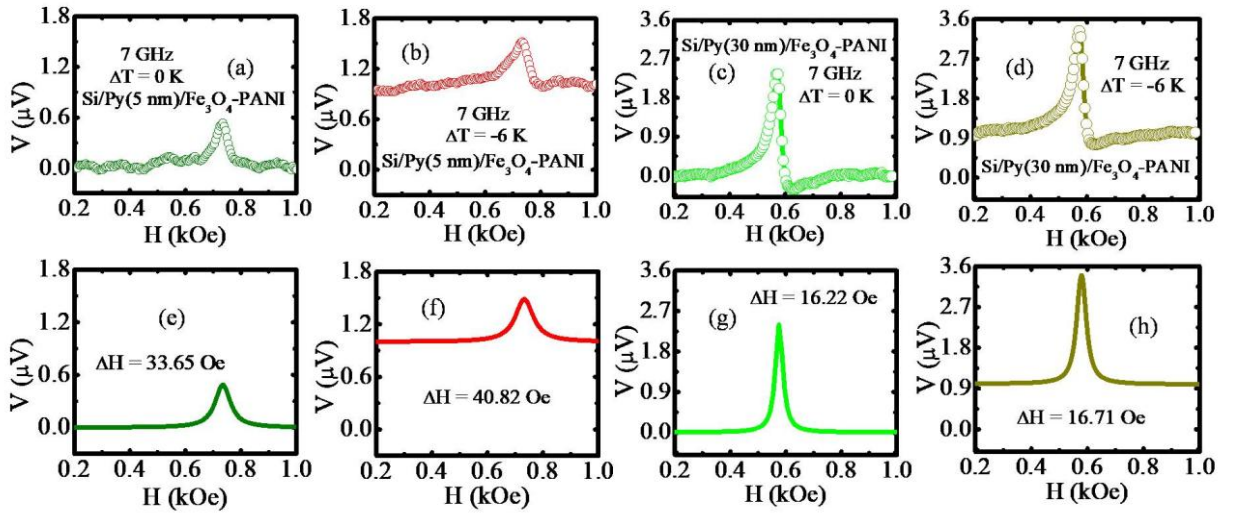

**Figure 2.** (color online) Shows the voltage V as a function of the applied magnetic field measured in the samples Si/Py(5 nm)/$Fe_3O_4$-PANI(200 nm) and Si/Py(30 nm)/$Fe_3O_4$-PANI(200 nm) at a frequency of 7 GHz, with a fixed power of $P_S$ = 20 mW. Panels **(a)** and **(c)** depict measurements without a temperature difference, while **(b)** and **(d)** show those with a temperature difference of ΔT = -6 K.

Panels **(e)**, **(f)**, **(g)**, and **(h)** illustrate the symmetric components from **(a)**, **(b)**, **(c)**, and **(d)**, displaying linewidths of ΔH = 33.65 Oe, ΔH = 40.82 Oe, ΔH = 16.22 Oe and ΔH = 16.71 Oe, respectively.

**Figure 3(a)** illustrates the Anomalous Nernst Effect (ANE) and Spin Seebeck Effect (SSE) charge currents, which were calculated by dividing the measured voltages by the resistances of the $Fe_3O_4$-PANI film over the length of the Peltier element (2 mm) as a function of ΔT, for H = ± 0.9 kOe and 7 GHz. In a metallic film such as Py, the application of a thermal gradient induces a voltage due to the ANE, while the spin current resulting from the SSE accumulates at the top of the Py film. Consequently, when microwave signals are applied, the spin current generated by the SSE is influenced by the SR and SP effects, which also convert into charge currents due to the Onsager reciprocity relations between different currents. **Figures 3(b)-(f)** present measurements of the field dependences of the voltage V, in the Si/Py/$Fe_3O_4$-PANI samples, resulting from currents generated simultaneously by SR, SP, SSE and ANE. These measurements were performed at frequencies of 5 and 7 GHz, with a temperature difference of ΔT = +6 K, where the Si substrate was cooler than the Si/Py/$Fe_3O_4$-PANI layers. Although the currents associated with the four processes (SR, SP, SSE, and ANE) differ in nature, they collectively produce a single voltage that represents the superposition of all effects.

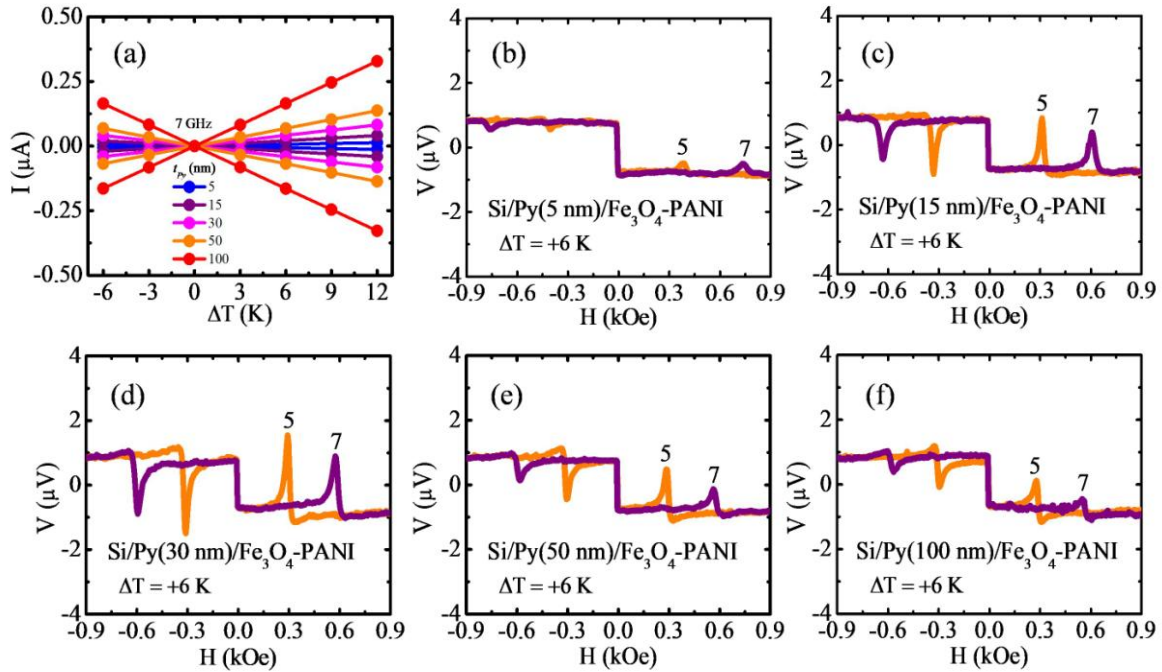

**Figure 3.** (color online) Illustrates **(a)** the ANE and SSE charge currents, which were calculated by dividing the measured voltages by the resistances of the Si/Py/$Fe_3O_4$-PANI bilayers over the length of the Peltier element (2 mm), as a function of ΔT for a magnetic field of ± 0.9 kOe and at a frequency of 7 GHz. Panels **(b)-(f)** show the measurements of the voltage (V) in the Si/Py/$Fe_3O_4$-PANI samples,

generated by currents simultaneously created by the SR, SP, SSE, and ANE effects, across frequencies from 5 to 7 GHz, with a temperature difference of ΔT = +6 K.

**Figures 3(b)-(f)** further demonstrate that the contributions of the SR and SP effects to the DC voltage, as a function of the thickness of the Si/Py/Fe$_3$O$_4$-PANI bilayers, exhibit two distinct regimes. The first regime corresponds to thicknesses ranging from 5 nm to 30 nm, where the voltages increase up to approximately 30 nm. This regime is characterized by the influence of the surface perpendicular anisotropy field, which is responsible for the symmetry breaking normal to the metallic ferromagnetic films. In the second regime, where the thickness exceeds 30 nm, the layers become sufficiently thick such that the perpendicular anisotropy field no longer plays a significant role. In this scenario, the contribution of the SP becomes small compared to that of the SR [9, 11]. The SR and SP effects are readily activated in Py films and are influenced by the spin Hall effect, and its inverse [11, 33, 34], but there have been no reports of thermal effects being a controlling factor.

**Figures 4(a)-(e)** display the variation of the linewidths with the temperature difference ΔT, measured in Si/Py($t_{Py}$)/Fe$_3$O$_4$-PANI at frequencies of 5, 6, 7, and 8 GHz. The linewidths were determined by fitting the data using equation (1). Despite the Si/Py/Fe$_3$O$_4$-PANI bilayer having a length of 4.5 mm and exhibiting nonuniform microwave excitation, only the uniform mode is excited, which is a commonly observed characteristic in Py samples [11-13, 17, 33, 34].

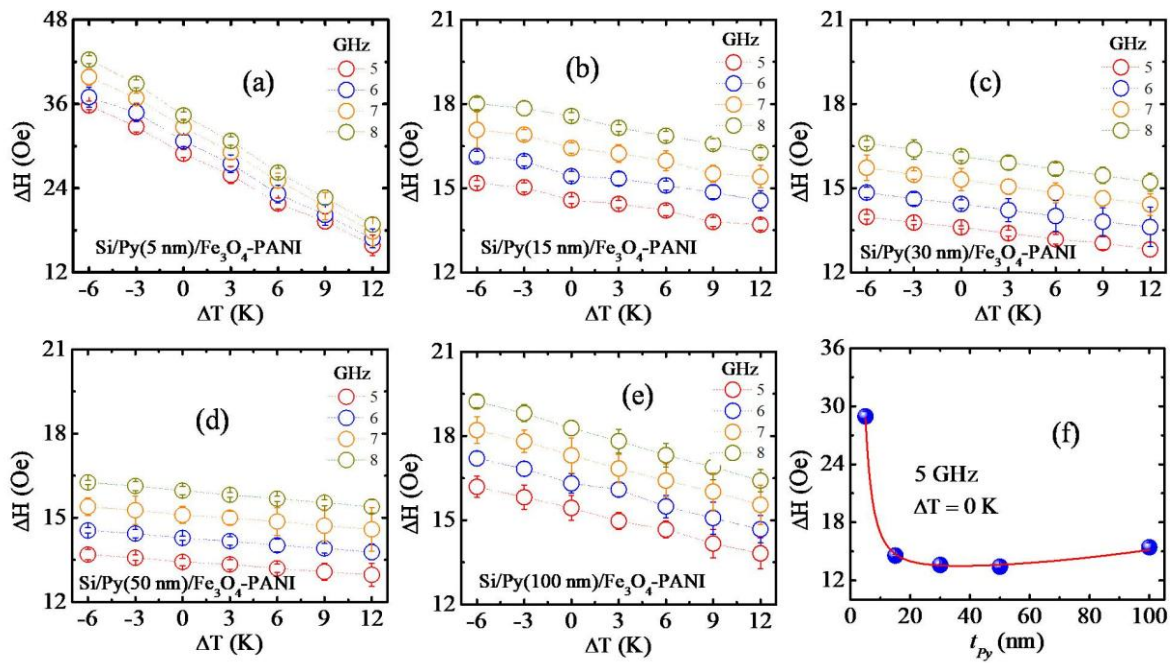

**Figure 4.** (color online) Presents **(a)-(e)** the linewidth as a function of the temperature difference measured in the Si/Py/Fe$_3$O$_4$-PANI bilayers at frequencies of 5, 6, 7, and 8 GHz. Panel **(f)** shows the dependence of the linewidth on the thickness of Si/Py($t_{Py}$)/Fe$_3$O$_4$-PANI bilayers for $\Delta T$ = 0 K and at a frequency of 5 GHz. A solid line is obtained using the expression for linewidth: $\Delta H = \Delta H_0 + \Delta H_{2M} + \Delta H_{Eddy} + \Delta H_{SP}$.

The results presented in **Figures 4(a)-(e)** indicate that the magnetic damping experiences a sharp decrease with thickness up to $t_{Py}$ = 30 nm, followed by a gradual increase at larger thicknesses, as illustrated in **Figure 4(f)**. The magnetic damping parameter, $\alpha = (\gamma/\omega)\Delta H$, significantly influences the behavior of the DC voltage linewidth, which is given by $\Delta H = \Delta H_0 + \Delta H_{2M} + \Delta H_{Eddy} + \Delta H_{SP}$. In this equation, $\Delta H_0$ represents the extrinsic contribution, $\Delta H_{2M}$ denotes the two-magnon scattering contribution and is calculated as $\Delta H_{2M} = C_{2M}/t_{Py}^2$ [35], while the eddy current contribution, $\Delta H_{Eddy} = C_{Eddy} t_{Py}^2$, becomes particularly important in the high-thickness regime of Py [36], and $\Delta H_{SP}$ is due to the inverse spin pumping effect. The solid line depicted in **Fig. 4(f)** was obtained with $\Delta H_0 + \Delta H_{SP} \approx 15$ Oe, $C_{2M}$ = 398 Oe × nm$^2$, and $C_{Eddy}$ = 2.5 × 10$^{-4}$ Oe/nm$^2$, which aligns well with previous studies [9, 11] for other bilayers. It is noteworthy that the absolute values of the linewidth changes shown in **Fig. 4** are over one order of magnitude larger than those observed in insulating ferrimagnets/heavy metal bilayers [3, 37, 38]. The ability to control magnetic damping is a fascinating phenomenon associated with spintronic thermal effects. The experiments conducted here reveal a decrease in damping with an increasing temperature gradient. This suggests that simultaneous measurements allow for the control of currents resulting from magnetization dynamics, influenced by thermal effects and governed by the Onsager reciprocity relations between spin and charge.

**CONCLUSIONS**

In summary, we measured the DC voltage (V) generated in Si/Py/Fe$_3$O$_4$-PANI bilayers under microwave-driven ferromagnetic resonance (FMR) conditions. This was conducted for a range of permalloy films with thicknesses varying from 5 to 100 nm while applying a thermal gradient. All DC voltage signals exhibited asymmetric line shapes, which are a combination of symmetric Lorentzian and antisymmetric Lorentzian derivative curves. The DC voltage measurements reflect the voltage resulting from the superposition of the SR, SP, SSE, and ANE effects. Our findings on the DC voltage linewidth indicate that the magnetic damping in Si/Py/Fe$_3$O$_4$-PANI bilayers can be controlled by currents generated due to the thermal gradient. The damping exhibited increases and decreases depending on the sign of the thermal gradient - a behavior that has never been reported before for a Si/Py/Fe$_3$O$_4$-PANI bilayer and is significantly larger than that observed in insulating ferrimagnets/heavy metal bilayers. Moreover, our results provide insights into the relationship between different phenomena in metallic materials, grounded in the Onsager reciprocity relations. This research marks a notable advancement in the field of spin calorimetrics.

**Acknowledgements**

This research was supported by Conselho Nacional de Desenvolvimento Científico e Tecnológico (CNPq) with Grant Number: 309982/2021-9, Coordenação de Aperfeiçoamento de Pessoal de Nível Superior (CAPES) with Grant Number: PROAP2024UFRPE, and Fundação de Amparo à Ciência e Tecnologia do Estado de Pernambuco (FACEPE) with Grant Number: APQ-1397-3.04/24. The authors are grateful to professor Francisco Estrada of Universidad Michoacana de San Nicolas de Hidalgo and professor Changjiang Liu of the Department of Physics of The State University of New York at Buffalo (University at Buffalo) for the valuable discussions on this research.

**Contributions**

J. L., C. E., L. P., and J. A. analyzed all the experimental measures and J. H. discussed, wrote and supervised the work.

**Conflict of interest**

The authors declare that they have no conflict of interest.

**Data availability statement**

Data will be made available on request.